\documentstyle[multicol,eqsecnum,aps,graphicx]{revtex}
\title{Electronic versus Phononic Friction of Xenon on Silver}
\author{A. Liebsch,$^1$ S. Gon\c calves,$^2$ and M. Kiwi$^{3,*}$}
\address{$^1$Institut f\"ur Festk\"orperforschung, Forschungszentrum
J\"ulich, 52425 J\"ulich, Germany\\
$^2$Instituto de F\'\i sica, Universidade Federal do Rio Grande do
Sul, Caixa Postal 15051, 91501-970 Porto Alegre (RS), Brasil\\
$^3$Facultad de F\'\i sica, Pontificia Universidad Cat\'olica,
Casilla 306, Santiago 22, Chile}

\begin{document}
\maketitle

\begin{abstract}
Molecular dynamics simulations of a Xe monolayer sliding on Ag(001)
and Ag(111) are carried out in order to ascertain the microscopic
origin of friction. For several values of the electronic contribution
to the friction of individual Xe atoms, the intra-overlayer phonon
dissipation is calculated as a function of the corrugation amplitude
of the substrate potential, which is a pertinent parameter to
consider. Within the accuracy of the numerical results and the
uncertainty with which the values of the relevant parameters are
known at present, we conclude that electronic and phononic dissipation
channels are of similar importance. While phonon friction gives rise
to the rapid variation with coverage, the electronic friction provides
a roughly coverage-independent contribution to the overall sliding
friction.
\end{abstract}

\begin{multicols}{2}

\section{Introduction} 
The microscopic origin of the sliding friction of thin films
adsorbed on a metal surface has recently attracted considerable
interest~\cite{bowden,persson}. The availability of refined
experimental methods~\cite{krim1,krim2,holzapfel} makes it now
feasible to investigate the fundamental processes that
contribute to the sliding friction on an atomic scale. On a
metal surface, frictional energy is dissipated via two main
channels: excitation of low-energy electron-hole pairs in the
substrate~\cite{SH81,Sols82,P91,PV95,sokoloff95,liebsch97,liebsch97b}
and excitation of phonons within the
overlayer~\cite{cieplak94,cieplak94b,PN96,TS97}. Phonon
excitation within the metal may also occur; however, this
contribution usually has only a minor influence on the lateral
sliding of adsorbed films.

Detailed experimental information on the friction of Xe and Kr
layers on noble metal surfaces was recently obtained by
Krim {\it et al.}~\cite{krim1,krim2} who performed quartz-crystal
microbalance (QCM) measurements for a variety of coverages. In this
technique, the frequency shift and broadening of the resonant line
provide information on the number of adsorbed atoms and on their
slippage during the lateral oscillation of the microbalance.
The data were analyzed by various theoretical groups that,
surprisingly, arrived at contradictory conclusions: For Kr on Au(111),
Cieplak {\it et al.}~\cite{cieplak94,cieplak94b} argued that the
observed friction
is caused by lattice vibrations within the overlayer and that
substrate-induced energy dissipation plays a negligible role. Dominant
phonon damping was also inferred for Xe on Ag(111)
by Tomassone {\it et al.}~\cite{TS97}.
On the other hand, Persson {\it et al.}~\cite{PN96} argued that the
phonon-associated friction for a full Xe monolayer on Ag(100) is
small and that the measured friction is mainly of electronic origin.
These opposite interpretations are puzzling since all three theoretical
studies were based on molecular dynamics simulations for
similar models of the rare gas/metal adsorption system. In addition,
Liebsch~\cite{liebsch97,liebsch97b} performed dynamical surface response
calculations within the time-dependent density functional approach
to determine the electronic friction of Xe atoms on Ag. The lateral
component of the friction coefficient was shown to agree qualitatively
with the experimental value for a full monolayer, suggesting that
phonon processes play a minor role in the compressed phase.

In view of these fundamental theoretical discrepancies, we
decided to carry out molecular dynamics simulations for Xe on
Ag, employing essentially the same overall model as previous
authors~\cite{cieplak94,cieplak94b,PN96,TS97}. Key input
parameters in these simulations are, in particular, the
corrugation amplitude of the Xe/Ag interaction potential and the
electronic friction of individual Xe atoms sliding parallel to
the substrate surface.  Since at present the corrugation
amplitude is known from independent experimental data only with
very appreciable uncertainty, a detailed understanding of how
its magnitude influences the final result is in our view
crucial. In contrast to the earlier molecular dynamics studies,
therefore, we do not choose a specific set of input model
parameters but investigate the net sliding friction within a
wide range of these parameters. In addition, we consider the
dependence on the crystallographic orientation by studying the
friction of a Xe layer adsorbed on Ag(111) and Ag(100), in order
to check the importance of the lateral symmetry of the substrate
potential. By carrying out such a systematic investigation we
essentially cover the full range of overlayer/substrate models
considered in the previous treatments.

A striking feature of the Xe/Ag quartz-crystal microbalance data by
Krim {\it et al.}~\cite{krim1,krim2} is the strong variation of the
net sliding friction with Xe coverage. To facilitate the discussion,
we show in Fig.~1 the measured slip time $\tau$ as a function
of Xe coverage~\cite{krim2} on Ag(111). The friction coefficient
$\bar\eta$ is given by the inverse of the slip time. According
to surface resistivity measurements for Xe on Ag~\cite{holzapfel},
the electronic friction coefficient depends weakly
on coverage. It is evident, therefore, that the rapid coverage
dependence observed in the QCM data, in particular near completion
of a monolayer, is caused primarily by
phonons generated within the overlayer. This is not surprising
since the probability of exciting phonons varies greatly with
inter-adatom spacing. With respect to this aspect of the data,
there is little dispute among the previous simulations studies.

\begin{figure}
\includegraphics[width=8.cm, clip=true]{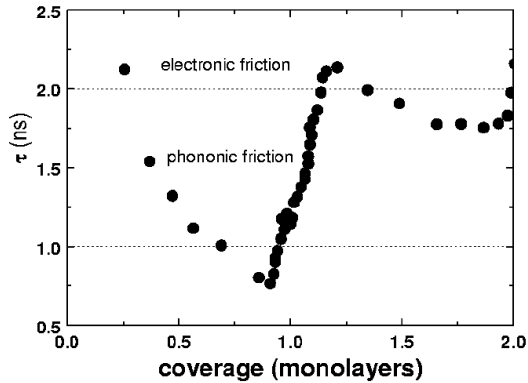}
\narrowtext
\caption{Measured slip time $\tau$ for Xe on Ag(111),
as a function of coverage, after Ref.~4.
In this paper we put forward arguments, based on extensive
simulations, to show that the region roughly between the
dotted lines can be understood as due to phonon originated friction,
while above the upper dotted line a friction mechanism of electronic
origin seems to prevail.}
\end{figure}

The important question, rather, concerns the relative weights of
electronic and phononic dissipation channels, especially at
coverages near a full monolayer where the slip time reaches its
maximum, i.e., where the sliding friction of the incommensurate
overlayer is smallest. Our results show that the effective
friction parameter $\bar\eta$ may be approximately written as
\begin{equation} 
\bar\eta = \eta_{el} + \eta_{ph}\ .
\end{equation} 
\noindent Here, $\eta_{el}=\eta_\Vert$ is the parallel component
of the single-atom electronic friction coefficient and the
intra-adsorbate phonon contribution is roughly given by
\,$\eta_{ph}=c\,u_0^2$, where $u_0$ is the corrugation amplitude
of the Xe/Ag potential.  A quadratic variation with $u_0$ agrees
with the analytical and numerical results of Cieplak {\it et
al.}~\cite{cieplak94,cieplak94b}.  The coefficient $c$ depends
weakly on the crystal face and on the single-atom friction
parameters $\eta_\Vert$ and $\eta_\perp$, but varies strongly
with coverage. Evidently, according to the above expression, the
measured sliding friction at any given coverage can be
reproduced using several combinations of the parameters $u_0$
and $\eta_{el}$. For instance, by reducing the single-site
electronic friction $\eta_\Vert$, one is able to match the data
if the substrate potential is assumed to be sufficiently
strongly corrugated.  Conversely, if the electronic friction is
chosen large enough, little additional phonon damping is
required to reproduce the same net $\bar\eta$.  Constraints on
the range of acceptable parameter values are implied by the
observed variation of the slip time with coverage, in
particular, by the steep slope near completion of the first
monolayer.

The molecular dynamics simulations for Xe on Ag, which we describe
below, suggest that the characteristic coverage dependence of the
observed slip time and its overall magnitude cannot be understood
only in terms of phonon dissipation. Although there is considerable
uncertainty both in the calculated and measured friction
coefficients, the most likely scenario is the presence of a roughly
coverage-independent electronic contribution of about \,$\tau\approx
2\,{\rm ns}$\, ($\eta_\Vert\approx0.5\,{\rm ns}^{-1}$) and an additional,
strongly coverage-dependent phonon friction, which can reduce the slip
time to about \,$\tau\approx1\,{\rm ns}$ (the net sliding friction then
increases approximately to \,$\bar\eta\approx1\,{\rm ns}^{-1}$).
Thus, roughly speaking, we associate the slip time in the region between
the horizontal dotted lines in Fig.~1 with phonon processes, while the
portion above the upper line is related to electronic energy dissipation.
Remarkably, the electronic friction deduced from this analysis is in
qualitative agreement with independent surface resistivity data for
Xe on Ag~\cite{holzapfel} and with theoretical estimates of this
dissipation channel~\cite{liebsch97,liebsch97b}. Thus, the friction of a full
monolayer is essentially of electronic origin, while the rapid
increase of the slip time before reaching the complete monolayer is
due to excitation of phonons within the Xe overlayer.

This paper is organized as follows: Section II summarizes the
theoretical results available for the electronic friction of Xe atoms
on a metal surface. Section III presents the model employed in the
molecular dynamics simulations. In particular, we specify the
interaction potentials and the friction felt by a single Xe atom.
Various technical aspects of the calculation of the
phonon friction are also discussed. Section IV contains the results,
with special emphasis on the variation of the net sliding friction
with the corrugation of the substrate potential and its dependence on
the electronic friction. Our conclusions are given in Section V.

\section{Electronic Friction}
We consider first the electronic contributions to the sliding friction
of individual Xe atoms above a metal surface. In a physisorption system,
these may arise due to the van der Waals attraction and the Pauli
repulsion. In the case of Xe, some dissipation might also be associated
with the formation of a weak covalent bond.

It is well known that the instantaneous mutual polarization between
neutral species gives rise to the long-range van der Waals attraction.
If an atom moves relative to a metal, the induced surface charge lags
behind, causing kinetic energy to be dissipated. Thus, as pointed out
long ago by various authors~\cite{SH81,Sols82}, the excitation of
low-frequency electron-hole pairs in the surface region contributes
to the friction of neutral atoms. If the adatom velocity is small
compared to the Fermi velocity of the metal electrons, the friction
coefficient may be expressed as \cite{Sols82,PV95}
\begin{equation}
    \eta_i = -{e^2\over m}\,\lim_{\omega\rightarrow0}\,{1\over\omega}
              \int\!d^3r\!\int\!d^3r' \, V_i({\bf r})
             \, {\rm Im}\,\chi({\bf r,r'},\omega)\, V_i({\bf r'}).
                                                        \label{etai}
\end{equation}
The index \,$i=x,z$\, refers to parallel or perpendicular motion of
the atom, \,$\chi$\, is the many-body response function of the metal,
and $m$ is the adatom mass. The effective interaction
\,$V_i({\bf r})$\, is approximately given by
\begin{equation}
         V_i({\bf r}) = {\alpha(0)\over 2}\,\nabla_i\left(\nabla
              {1\over\vert{\bf r - d}\vert}\right)^2 \ .\label{Vi}
\end{equation}
Here, \,$\alpha(0)$\, is the static polarizability of the adatom and
  \,${\bf d}=(0,0,d)$\, is the location relative to the jellium edge.
Since \,$V_i({\bf r})$\, does not satisfy the Laplace equation,
it can only be approximately represented in terms of a superposition
of evanescent plane waves of the form \cite{PV95}
\begin{equation}
   V_i({\bf r}) =    \sum_{\bf q_\Vert}\ V_i({\bf  q_\Vert})
               \ e^{i{\bf q_\Vert}\cdot{\bf r_\Vert} + qz} \ .
\end{equation}
With this expansion we can write $\eta_i$ as
\begin{equation}
      \eta_i={e^2\over m}\ {1\over2\pi }\ \lim_{\omega\rightarrow0}\,
         {1\over\omega}
        \sum_{\bf q_\Vert}\ \vert V_i({\bf q_\Vert})\vert^2
           \ q \ {\rm Im}\,g(q,\omega) \ ,     \label{img}
\end{equation}
where the surface response function $g(q,\omega)$ is defined as
\begin{equation}
      g(q,\omega) = \int dz\ e^{qz}\ n_1(z,q,\omega)
\end{equation}
and \,$n_1(z,q,\omega)$\, is the electronic surface density induced
by an external potential given by
\,$-{2\pi\over q}\,e^{i\vec q_\vert\cdot\vec r_\Vert +qz}$.

In the case of Xe atoms physisorbed on Ag(111) at a temperature
$T=77.4$~K, the distance $d=2.4$~\AA.  The static polarizability
of Xe is \,$\alpha(0)=4.0$~\AA$^3$. Using the density functional
results for \,$n_1(z,q,\omega)$\, and \,$g(q,\omega)$, we find
\,$\eta_\Vert\sim 0.34$~ns$^{-1}$\, and
\,$\eta_\perp\sim1.74$~ns$^{-1}$ \cite{liebsch97,liebsch97b}.

We note that the theoretical value of $\eta_\Vert$ given above
presumably represents a slight overestimate since the Xe atom is
not completely outside the range of the electronic density profile
as implicitely assumed in (\ref{img}). Moreover, as a result of
the hybridization between the Ag $s$ and $d$ electrons, the surface
polarizability of real Ag may be slightly smaller than that of
the corresponding jellium model~\cite{inglesfield}. We also point
out that, according to (\ref{Vi}), the friction coefficient varies
asymptotically like \,$1/d^{10}$, i.e., it is rather sensitive to
the precise location of the adatom above the metal surface.

The friction coefficient arising from the van der Waals
attraction derived above is significantly larger than the
contribution due to the Pauli repulsion which was estimated by
Persson~\cite{P91} to be about \,$\eta_\Vert\sim
0.06$~ns$^{-1}$. In addition, for Xe there may exist some
friction due to chemical effects resulting from the broadening
of the Xe $6s$ level. Estimates of this mechanism
yield~\cite{P91} \,$\eta_\Vert\sim 0.15$~ns$^{-1}$.

On the basis of these theoretical calculations and estimates, we
conclude that the total electronic contribution to the lateral
friction of single Xe atoms on Ag is approximately
\,$0.5\ldots0.6$~ns$^{-1}$. At full monolayer coverage, this
friction is presumably slightly smaller because of the weakening of
the Xe/Ag bonds. An electronic contribution of this magnitude is in
excellent agreement with the surface resistivity measurements for
Xe on Ag~\cite{holzapfel}.

According to our molecular dynamics simulations at $T=77.4$~K,
the distance of Xe atoms normal to the surface varies about
0.1~\AA\ around the equilibrium distance \,$d=2.4$~\AA.  For the
asymptotic van der Waals attraction, such a variation implies a
variation of $\eta_\Vert$ roughly between 0.2 and 0.5~ns$^{-1}$.
For the Pauli and covalent-bond contributions the variation is
exponential due to wave function overlap, with an exponent
determined by the work function. This leads to a variation of
$\eta_\Vert$ of about 20 \%. Thus, for some Xe atoms, the
electronic friction is larger, for others smaller than the
equilibrium value \,$0.5\ldots0.6$~ns$^{-1}$\, quoted above.
Because of this uncertainty the simulations are carried out for
a sufficiently wide range of electronic friction coefficients.

\section{Model for Phonon Friction}
In this section we focus on the evaluation of the friction induced by
the excitation of phonons within the adsorbed layer of Xe atoms. The
single-atom frictional properties are assumed to be known and are
used as inputs in the molecular dynamics simulation. We specify first
the intra-adsorbate and adsorbate--substrate potentials and then discuss
details concerning the molecular dynamics simulations.

\subsection{Potentials}
As in previous work~\cite{cieplak94,cieplak94b,PN96,TS97}, the interaction
between Xe atoms is expressed as a sum of Lennard-Jones pair
potentials
\begin{equation}
    v(r) = \epsilon\,\left[\left(r_0\over r\right)^{12} -
                          2\left(r_0\over r\right)^6 \right]\ ,
\end{equation}
\noindent
where $\epsilon=19$ meV is the well depth and $r_0=4.54$ \AA\ is
the inter-particle spacing at the potential minimum.
The total Xe-Xe interaction potential then takes the form
\begin{equation}
    V = {1\over2}\sum_{i\neq j} v({\bf r}_i -{\bf r}_j)\ .
\end{equation}
\noindent
The adsorbate-substrate interaction potential is assumed to be
given by
\begin{eqnarray}
   U(\bf r) &=& E_0\,\left[ e^{-2\alpha(z-z_0)}
                        - 2 e^{- \alpha(z-z_0)} \right] \nonumber\\
            & & + \ u_0\, e^{-\alpha'(z-z_0)}\, u(x,y) \ , \label{U}
\end{eqnarray}
\noindent
where for Ag(100)
\begin{equation}
     u(x,y) = 2 + \cos(kx) + \cos(ky) \ ,
\end{equation}
\noindent
whereas, in the case of Ag(111),
\begin{eqnarray}
     u(x,y) &=&  1.5 +\cos\Big(2ky/\sqrt3\Big)
                     +\cos\Big(k(x+y/\sqrt3)\Big)\nonumber\\
            &&       +\cos\Big(k(x-y/\sqrt3)\Big)\ .
\end{eqnarray}
\noindent
Here, $k=2\pi/a$ and $a=2.89$ \AA\ is the spacing between neighboring
Ag atoms.  The unit vectors of the substrate lattice are $(a,0), \ (0,a)$
for Ag(100) and $(a,0), \ (a/2,a\sqrt3/2)$ for Ag(111). The binding
energy for Xe on Ag is $E_0=0.23$~eV. The decay constant $\alpha$ may
be deduced from the frequency of the perpendicular vibration~\cite{PN96},
i.e., $\alpha=0.72$~\AA$^{-1}$. The decay constant $\alpha'$ of the
Fourier component should in principle be larger than $\alpha$.
However, this difference is not very important and we choose
$\alpha'=\alpha$.  The corrugation factor $u(x,y)$ is assumed to
vanish at hollow adsorption sites. The potential barrier between
neighboring hollow sites is $2u_0$ for the (100) face and $0.5u_0$
for the (111) face. The total barrier between hollow sites across
the substrate lattice sites is $4u_0$ for the (100) face and $4.5u_0$
for (111).

The adsorbate--substrate interaction specified in Eq.~(\ref{U}) has
the same form as the one used by Persson and Nitzan~\cite{PN96}.
Cieplak {\it et al.}~\cite{cieplak94,cieplak94b} and Tomassone {\it et
al.}~\cite{TS97}, on the other hand, used the expression derived by
Steele~\cite{steele} for a sum of pair potentials. Since this
potential is much more corrugated than the true interaction between
rare gas atoms and metal surfaces, the coefficient of the
non-vanishing Fourier component was drastically scaled down.
The functional form of this potential also differs from expression
(\ref{U}). However, the precise shape near the potential minimum
should not be crucial, as long as the overall corrugation is the
same.

Since the parameter $u_0$ is decisive for the excitation of phonons
within the overlayer, we do not choose a particular value. Instead, we
calculate the net sliding friction as a function of $u_0$ in order to
illustrate its sensitivity to the adsorbate--substrate potential.
In this manner, we cover the various substrate potentials considered
previously.

\subsection{Simulations} There are several methods of extracting
the friction parameter from the molecular dynamics simulations.
(i)~One can apply a constant lateral force $\bf F$ to all the
adatoms, while keeping the substrate atoms at rest.  After an
appropriate number of time steps, the steady-state velocity $\bf
v$ of the overlayer is determined. If the friction is viscous,
$\bf v$ satisfies the linear relation \begin{equation} {\bf F} =
m\,\bar\eta \,{\bf v}\ , \end{equation} where $m$ is the adatom
mass. This procedure was used by Cieplak {\it et
al.}~\cite{cieplak94,cieplak94b} and by Persson and Nitzan~\cite{PN96};\
(ii)~To simulate the quartz-crystal microbalance measurement,
one may let the substrate atoms oscillate laterally and
determine the induced lateral vibration of the adatoms. This
method was also used by Cieplak {\it et al.}~\cite{cieplak94,cieplak94b};\
(iii)~One can apply a constant force to the adatoms up to a
specific point in time and then switch it off. If the friction
is viscous, the decay of the steady-state velocity of the
adsorbate is proportional to $e^{-\bar\eta t}$;\ (iv)~Finally,
one may determine $\bar\eta$ from the thermal equilibrium
autocorrelation function of the center of mass velocity of the
overlayer.  The latter two methods were employed by Tomassone
{\it et al.}~\cite{TS97}.

In the present work, we also apply a constant lateral force $\bf F$
to the adatoms and determine the steady-state velocity that is reached
after a sufficiently long time. The simulations are based
on the Langevin equation
\begin{equation}
   m\ddot{\bf r}_i+m\eta\dot{\bf r}_i = - {\partial U\over\partial{\bf r}_i}
                                        - {\partial V\over\partial{\bf r}_i}
                                        + {\bf f}_i + {\bf F} \ ,\label{LE}
\end{equation}
\noindent
where ${\bf f}_i$ is a stochastically fluctuating force describing
the effect of the irregular thermal motion of the substrate on the
$i^{th}$ adatom. The components of ${\bf f}_i$ are related to the
microscopic friction $\eta$ via the fluctuation-dissipation theorem
\begin{equation} \langle f_i^\alpha(t)\,f_j^\beta(0) \rangle = 2mk_{\rm B} T
\eta_{\alpha\beta} \delta_{ij}\delta(t)\ .
\end{equation}
\noindent
The microscopic friction tensor $\eta_{\alpha\beta}$ is assumed to be
diagonal, with independent elements $\eta_\Vert$ and $\eta_\perp$.

To solve Eq.~(\ref{LE}) the time variable is discretized with a
step duration $0.01\,t_0$, where the natural time unit is
\,$t_0=r_0\,(m/\epsilon)^{1/2}= 3.84$~ps. The simulations are
carried out using the integration procedure suggested by Tully
{\it et al.}~\cite{tully}.  Typically, thermalization was
reached after $10^2-10^3\,t_0$.  The drift friction $\bar\eta$
was derived using the definition \,$\bar\eta=F/(m\langle
v\rangle)$, where the external force $F$ was applied in the
$x$-direction on each adsorbate, resulting in the (time
averaged) drift velocity $\langle v\rangle$.  On Ag(100), the
basic unit was taken as a square containing $12\times12$ or
$24\times24$ substrate atoms. Similar unit cells were employed
for Ag(111).  These cells were then repeated assuming periodic
boundary conditions.

Except for the form of the substrate potential, the model
outlined above is very similar to the ones used in earlier
simulations~\cite{cieplak94,cieplak94b,PN96,TS97}. The main difference
between the treatments Of Refs.~\cite{cieplak94,cieplak94b,PN96,TS97}
resides in the interpretation of the microscopic friction of a
single Xe adatom. In the work by Cieplak {\it et
al.}~\cite{cieplak94,cieplak94b} this friction plays the role of a
thermostat that allows to establish a constant temperature $T$
within the overlayer. Since the direction of this single-site
friction is chosen orthogonal to the direction of the external
force, it was claimed not to affect the net sliding friction of
the overlayer. Instead, in the work by Persson and
Nitzan~\cite{PN96} this thermostat was given a specific
microscopic origin.  In particular, $\eta_\Vert$ coincides with
the lateral friction of individual Xe atoms due to excitation of
electron-hole pairs in the substrate, i.e.,
$\eta_\Vert=\eta_{el}$. According to the theoretical estimates
and the resistivity measurements discussed in Section II, this
coefficient should be about \,$\eta_\Vert\approx 0.5$~ns$^{-1}$.
Finally, in the work by Tomassone {\it et al.}~\cite{TS97}, the
simulations were carried out in the absence of a thermostat or
any other single-site friction forces.

In principle, electronic processes also contribute to the friction of
the perpendicular motion of single adatoms. As pointed out in Section
II, density functional calculations yield
\,$\eta_\perp\approx1.74$~ns$^{-1}$.
However, this contribution is much smaller than the normal friction
induced by phonon excitation, which was estimated by Persson and
Nitzan~\cite{PN96} to be roughly \,$\eta_\perp=260$~ns$^{-1}$.
Because of the present uncertainties in the evaluation of both
$\eta_\Vert$ and $\eta_\perp$, we have carried out simulations for
a whole range of these parameters, in order to illustrate the
sensitivity of the net sliding friction to the microscopic
processes governing the behavior of individual Xe atoms.

\section{Results and Discussion} In this section we first
discuss the sliding friction of a monolayer of Xe atoms on Ag
and then address the coverage dependence. In the final part, we
compare these results with the QCM data. The temperature of
\,$T=77.4$~K, which we adopt, corresponds to the one used in the
measurements by Krim {\it et al.}~\cite{krim1}. We define the
coverage \,$\Theta=1$\, as the fully compressed
monolayer~\cite{dai94} with density $n_a=0.0597$~\AA$^{-2}$. On
Ag(100) this amounts to \,$N=72$\, Xe atoms within the basic
\,$12a\times12a$\, unit cell ($N=288$ within the $24a\times24a$
cell). The uncompressed monolayer has a slightly lower density,
\,$n_a=0.0565$~\AA$^{-2}$, i.e., \,$\Theta=0.94$ ($N=68$\, atoms
within the \,$12a\times12a$\, unit cell).  The remaining
parameters required in the simulations, which we report below,
have been specified in the preceding sections.

\subsection{Corrugation dependence of the sliding friction for
an uncompressed Xe layer}
In Fig.~2 we display typical results of the center of mass
velocity, in the direction of the applied force, as a function of
time. The examples are for $N=68$ on the (100) surface and the
constant external force on each of the adsorbates \,$F=0.001\,
\epsilon/r_0$\, is applied after \,$100\,t_0$\, as indicated by
the arrow. The single-atom friction coefficients are
\,$\eta_\Vert= 0$\, and \,$1.3$~ns$^{-1}$, with
\,$\eta_\perp=260$~ns$^{-1}$.  The corrugation amplitude is
\,$u_0=0.95$~meV and $1.9$~meV.  

\begin{figure}
\includegraphics[width=8.cm, clip=true]{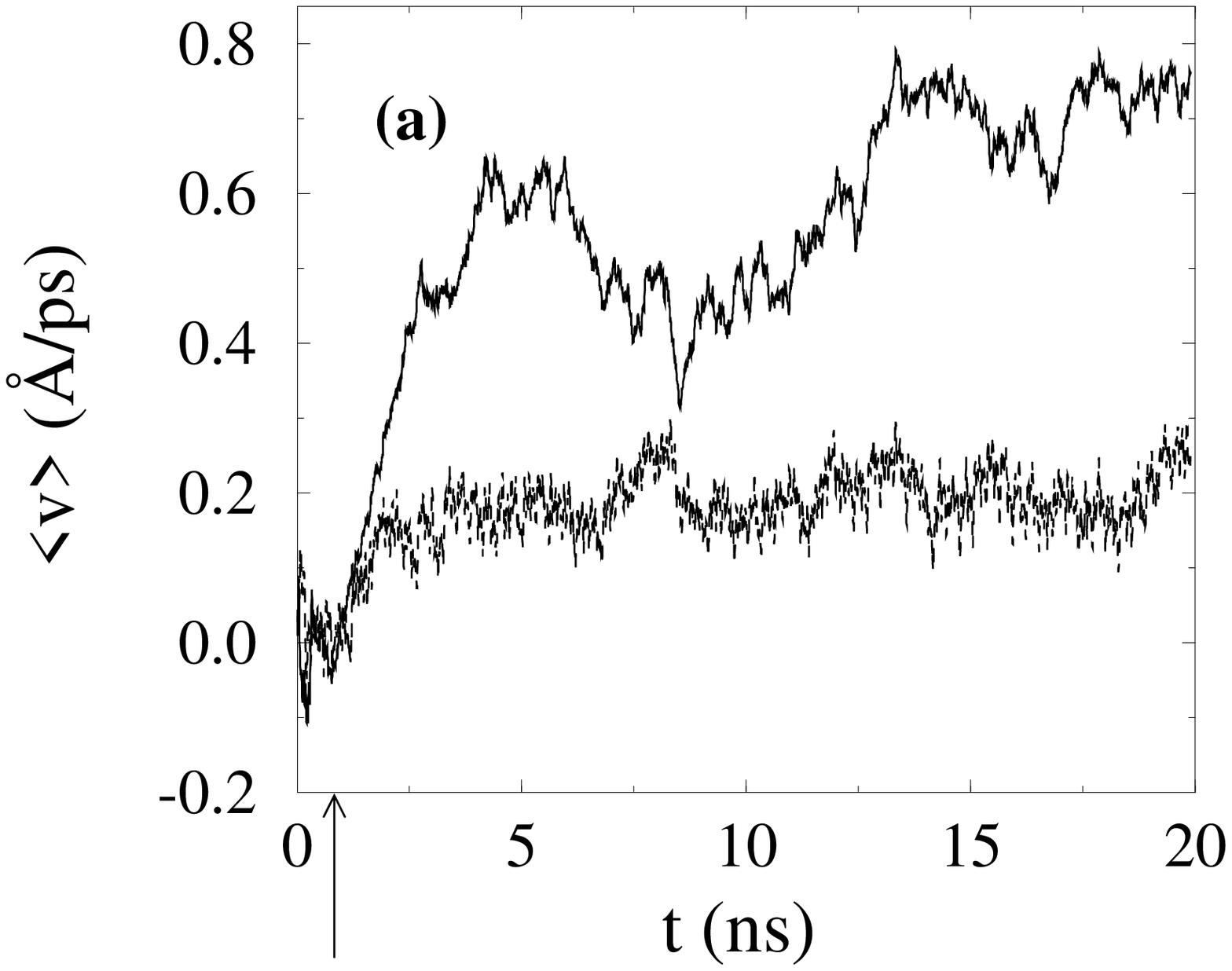}

\includegraphics[width=8.cm, clip=true]{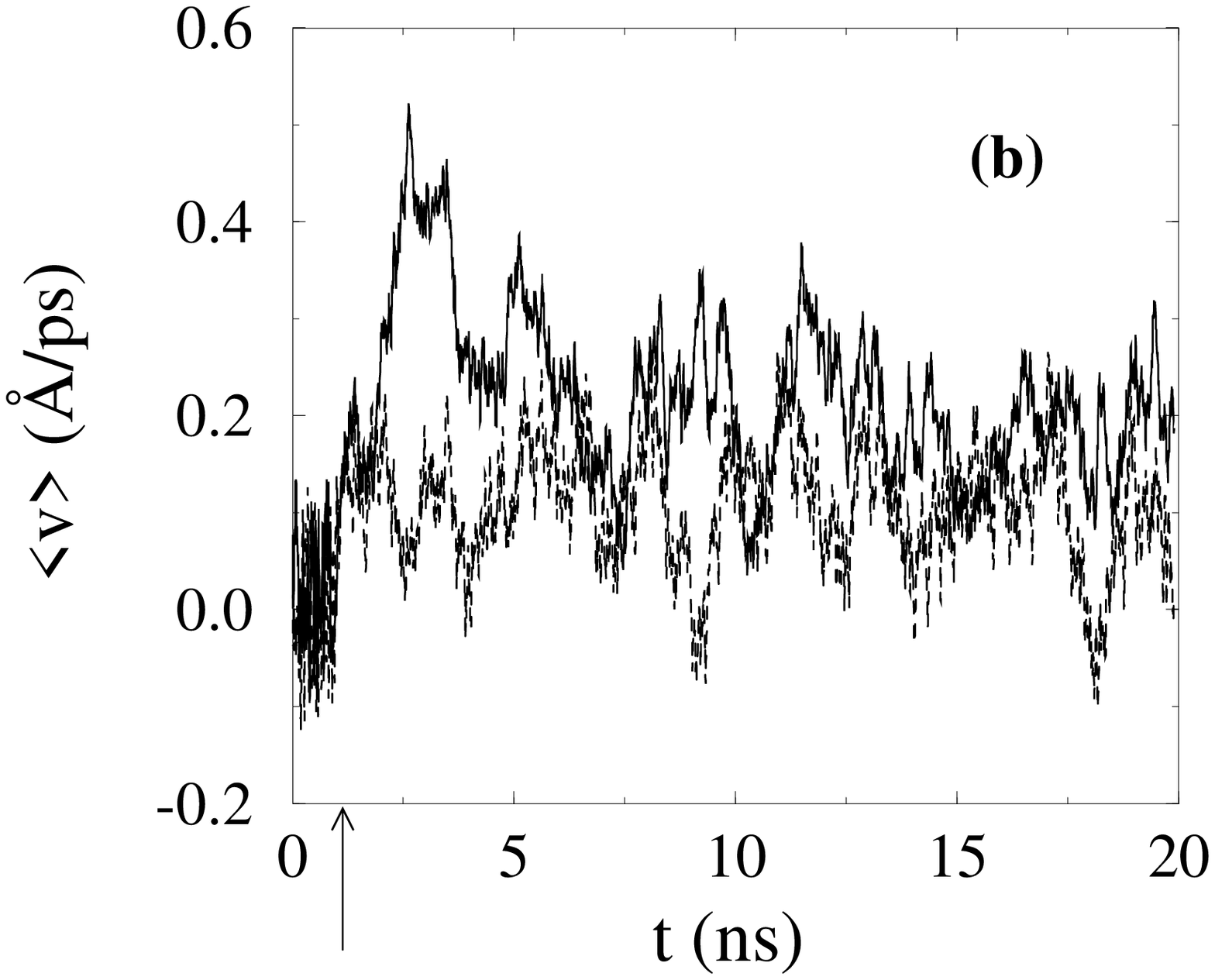}
\narrowtext
\caption{Center of mass velocity (in units \AA/ps)
of a Xe layer on Ag(100) as a function of time. The coverage
corresponds to an uncompressed monolayer at $\Theta = 0.94$ or
$n_a = 0.0565$ atoms/\AA$^2$. The arrow indicates the time
($t=100\,t_0=384$~ps) at which the external force
\,$F=0.001\,\epsilon/r_0=4.2\times 10^{-3}$~meV/\AA\
is applied. Upper curves are for $\eta_\Vert=0$, lower curves
correspond to \,$\eta_\Vert=1.3$~ns$^{-1}$. (a)~$u_0=0.95$~meV;
(b)~$u_0=1.9$~meV.}
\end{figure}

One of the conclusions of the present work can be derived 
directly from this figure. 
In Fig.~2a, for a relatively low corrugation amplitude
\,$u_0=0.95$~meV, the effect of the single-site friction
coefficient $\eta_\Vert$ is evident: suppressing it, the center
of mass velocity increases and so do the fluctuations. But the
effect of the suppression is attenuated when the corrugation
amplitude is bigger, as is the case in Fig.~2b for
\,$u_0=1.9$~meV. Thus, the relative importance of phononic or
electronic dissipation channels depends crucially on the
magnitudes of $u_0$ and $\eta_\Vert$.

The validity of the linear response regime can be checked in Fig.~3, where
we show the center of mass velocity $\langle v\rangle$, time averaged over
\,$4000\,t_0$\, after the transient period following the application
of the external force $\bf F$, as a function of $F$. The approximate
linear relation \,$F = m\bar\eta\langle v\rangle$\, is apparent in all
cases, indicating that the system indeed obeys a viscous force law.
The full lines represent weighted linear regressions of the
\,$\langle v\rangle\sim F$\, relation, which we use to derive the net
sliding friction $\bar\eta$.

\begin{figure}
\includegraphics[width=7.cm, clip=true]{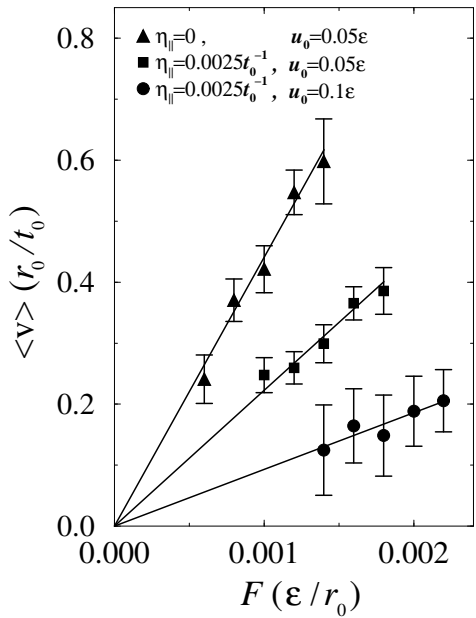}
\narrowtext
\caption{Average center of mass velocity $\langle v\rangle$
(in units of $r_0/t_0$) as a function of external force $F$
(in units of $\epsilon/r_0$) showing the linear viscous regime
for different values of the input parameters $\eta_\Vert$ and $u_0$.
Each point was obtained by time
averaging over a long steady state run and over two independent runs.
Error bars were estimated from the fluctuations of the center of mass
velocity as can be seen in Fig.~2. Straight lines are linear fits
of points accounting for error bars. The values of $\bar\eta$
are as follows (from top to bottom):
\,$0.0023/t_0$, \,$0.0045/t_0$, and \,$0.010/t_0$.}
\end{figure}

Fig.~4 summarizes our results for $\bar\eta$ as a function of the
corrugation amplitude $u_0$ and for different values of the lateral
electronic single-atom friction $\eta_\Vert$.
The overall sliding friction is approximately of the form
\begin{equation}
     \bar\eta = \eta_{el} +  c\, u_0^2\ ,    \label{c}
\end{equation}
\noindent
with \,$\eta_{el}=\eta_\Vert$. The intra-adsorbate phonon contribution
\,$\eta_{ph}=c\,u_0^2$\, varies quadratically with the corrugation
amplitude $u_0$ and the coefficient $c$ depends weakly on the
coefficient $\eta_\Vert$ and on the crystal structure of the substrate.
This can be seen by comparing the results shown in Fig.~4a and 4b for
the Ag(100) and Ag(111) surfaces, respectively. Although the overall
variation of $\bar\eta$ is similar for both faces, the coefficient $c$
is slightly larger for Ag(111) than for Ag(100), indicating that $\bar\eta$
is governed by the total barrier height ($4.5\,u_0$\, for (111) vs.
$4\,u_0$\, for (100)), rather than by the corrugation barrier between
neighboring hollow sites ($0.5\,u_0$\, for (111) vs. $2\,u_0$\, for
(100); see previous section). This trend is plausible since, at
monolayer coverage, the Xe atoms are not free to move along paths
between nearest potential minima but are forced across potential maxima.
We note here that, in the  \,$\eta_\Vert=0$ case, a quadratic variation of
$\bar\eta$ with $u_0$ was also found by Cieplak {\it et
al.}~\cite{cieplak94,cieplak94b}.

\begin{figure}
\includegraphics[width=7.cm, clip=true]{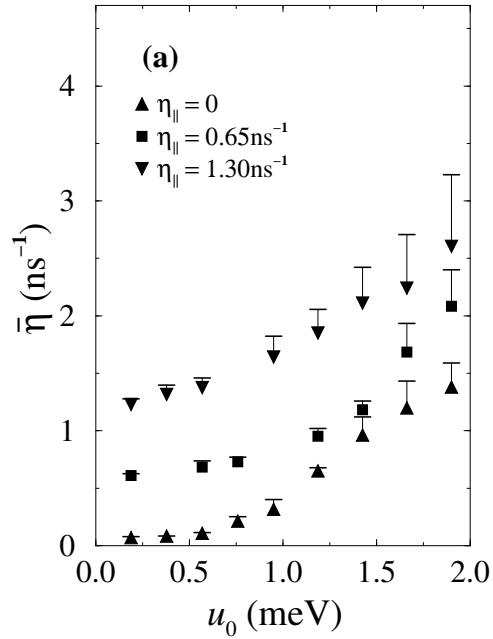}

\includegraphics[width=7.cm, clip=true]{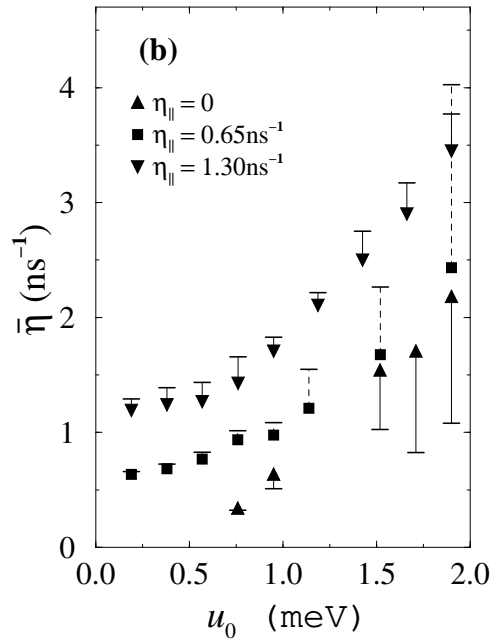}
\narrowtext
\caption{Net sliding friction of Xe layer on Ag as
a function of corrugation amplitude $u_0$. The coverage is
$\Theta=0.94$. The results are shown for three values of the lateral
single-atom electronic friction, indicated by the symbols:
\,$\eta_\Vert=\eta_{el}=0$, \,$0.0025\,/t_0$, and \,$0.0050\,/t_0$.
\,$\eta_\perp=1.0\,/t_0$ in all cases. (a) (100) surface; (b) (111) surface.}
\end{figure}

In order to demonstrate the weak dependence of the coefficient $c$ on
$\eta_\Vert$, we collapse in Fig.~5 the data obtained for different
$\eta_\Vert$ into one curve by plotting
\begin{equation}
      \eta_{ph} = \bar\eta - \eta_\Vert\ .
\end{equation}
The full lines correspond to a quadratic fit of the data points.
According to Figs.~5a and 5b, we deduce for the (100) and (111) faces
\,$c\approx0.6\,t_0^{-1}\epsilon^{-2}\approx 0.42$~ns$^{-1}$meV$^{-2}$, and
\,$c\approx0.8\,t_0^{-1}\epsilon^{-2}\approx 0.56$~ns$^{-1}$meV$^{-2}$,
respectively.

\begin{figure}
\includegraphics[width=7.cm, clip=true]{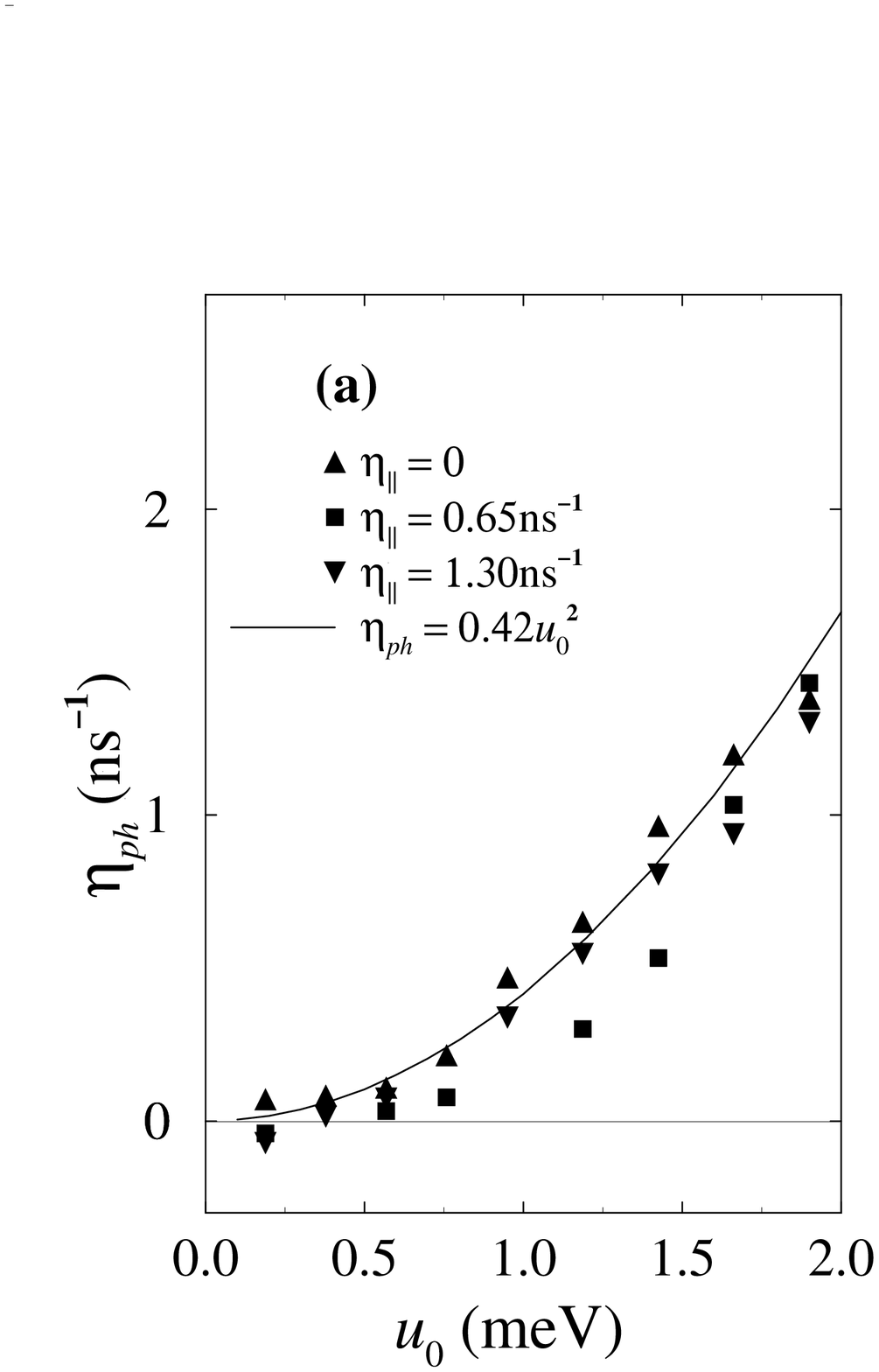}

\includegraphics[width=7.cm, clip=true]{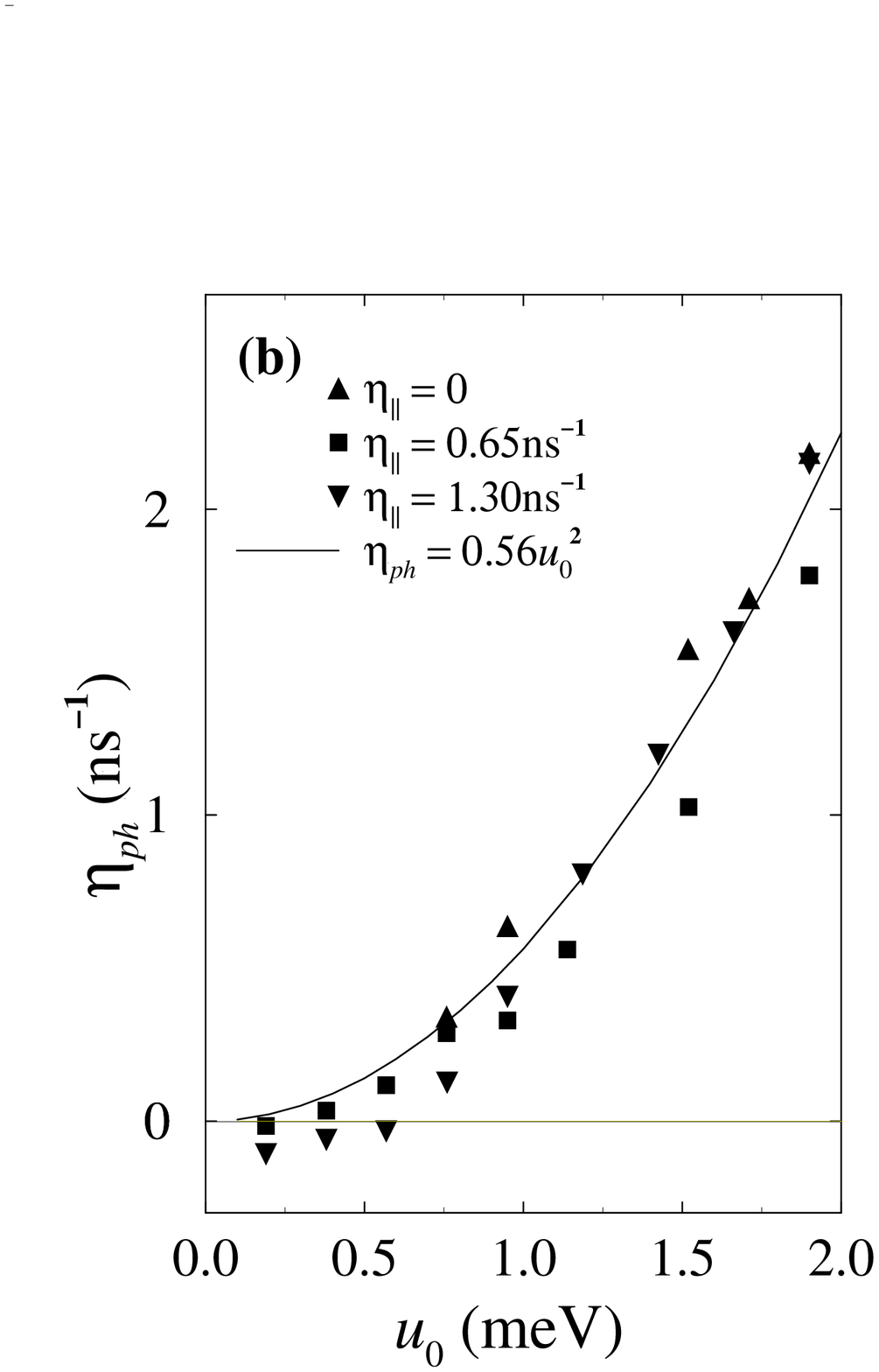}
\narrowtext
\caption{Phonon contribution to the sliding friction of
Xe layer on Ag obtained as $\bar\eta - \eta_\Vert$, as a function of
corrugation amplitude $u_0$. Symbols as in Fig.~4. Full
lines are quadratic fits accounting for errors of points. Error bars
are the same as Fig.~4 but are suppressed here for clarity.
(a) (100) surface; (b) (111) surface.}
\end{figure}

The results shown in Figs.~4 and 5 correspond to
\,$\eta_\perp=1/t_0 = 260$~ns$^{-1}$.  Simulations for different
values of $\eta_\perp$ in the range
\,$26$~ns$^{-1}\leq\eta_\perp\leq 260$~ns$^{-1}$\, yield nearly
the same overall sliding friction, confirming the results of
previous studies which suggested that the friction in the
direction of the external force is essentially independent of
the thermostat or the single-site friction in the orthogonal
direction~\cite{cieplak94,cieplak94b,PN96}.

The simulations discussed so far were carried out for 68 Xe atoms
within substrate unit cells corresponding to \,$12a\times12a$\, for
Ag(100) and \,$12a\times14a\sqrt3/2$\, for Ag(111). Remarkably, we
observed pronounced finite-size effects associated with the small
dimension of these unit cells. This is illustrated for Ag(100)
in Fig.~6, where $\eta_{ph}$ is plotted as a function of $u_0$.
In general the Xe atoms form approximate hexagonal
layers. Their orientation, however, tends to prefer two angles:
with rows of Xe atoms nearly parallel to the $x$-axis, or nearly
parallel to the $y$-axis. We denote these phases by $\alpha$ and $\beta$,
respectively. Since these orientations persist even in the absence
of the substrate corrugation ($u_0=0$) and for vanishing external
force, it is clear that they are a consequence of the limited size
of the unit cell. The reason is that it is not possible to accomodate
truly incommensurate hexagonal overlayers, at arbitrary angular
orientations, in a rectangular unit cell that is periodically repeated.

\begin{figure}
\includegraphics[width=7.cm, clip=true]{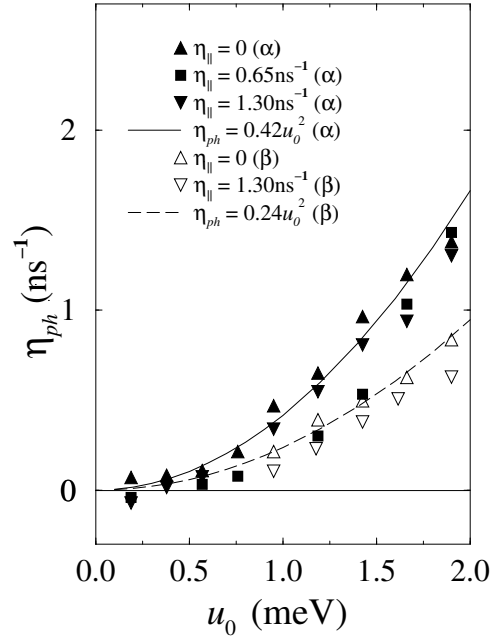}
\narrowtext
\caption{Phonon contribution to the sliding friction of
Xe layer on Ag(100) obtained as $\bar\eta - \eta_\Vert$, as a function of
corrugation amplitude $u_0$. Symbols and quadratic fits as in Fig.~5.
Full symbols and full line: $\alpha$-phase, open symbols and dashed
line: $\beta$-phase, corresponding to different angular orientation
of Xe monolayer.}
\end{figure}

In order to reduce these finite-size effects, we have performed
simulations for much larger substrate cells (e.g., 288 Xe atoms
in a \, $24a\times24a$\, cell on Ag(100)). Qualitatively, the
net sliding friction for these systems agrees rather well with
the earlier results, but the tendency for orientational alignment
of the Xe layer is greatly reduced.

\subsection{Coverage dependence of sliding friction}
As mentioned above, the coefficient $c$ of the quadratic term in
Eq.~(\ref{c}) depends strongly on overlayer coverage. We illustrate
this point in Fig.~7, where the net sliding friction of the uncompressed
monolayer ($\Theta=0.94$) is compared with that for a fully compressed
Xe layer ($\Theta=1.0$) and for a less dense layer ($\Theta=0.85$).
The compression is seen to lead to a greatly reduced phonon-induced
contribution to $\bar\eta$. On Ag(100)
as well as Ag(111), the magnitude of $c$ is reduced by more than a
factor of two. A strong reduction of dissipation via phonons upon
compression is to be expected since the Xe atoms are much less free
to vibrate about the equilibrium positions.
The opposite trend is found if the coverage is reduced below that of
the uncompressed monolayer. For Xe on Ag(100) at coverage $\Theta=0.85$,
the coefficient $c$ is about \,$1.25$~ns$^{-1}$meV$^{-2}$, i.e., three
times larger than for $\Theta=0.94$. An even more dramatic increase
of $c$ is found on Ag(111). The strong enhancement of $\bar\eta$ is
caused by the higher probability of exciting atomic vibrations in
the less dense overlayer phase.

\begin{figure}
\includegraphics[width=7.cm, clip=true]{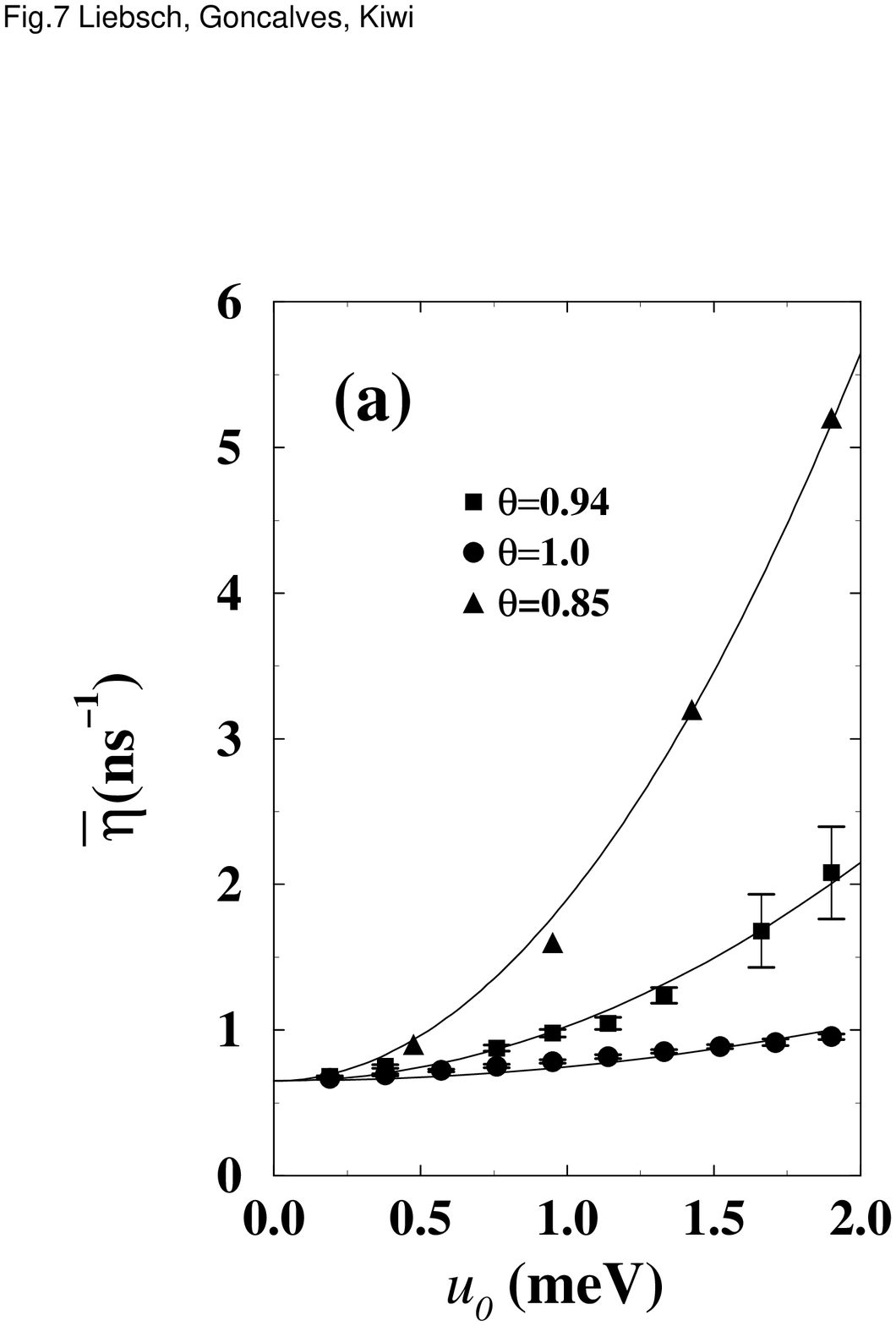}

\includegraphics[width=7.cm, clip=true]{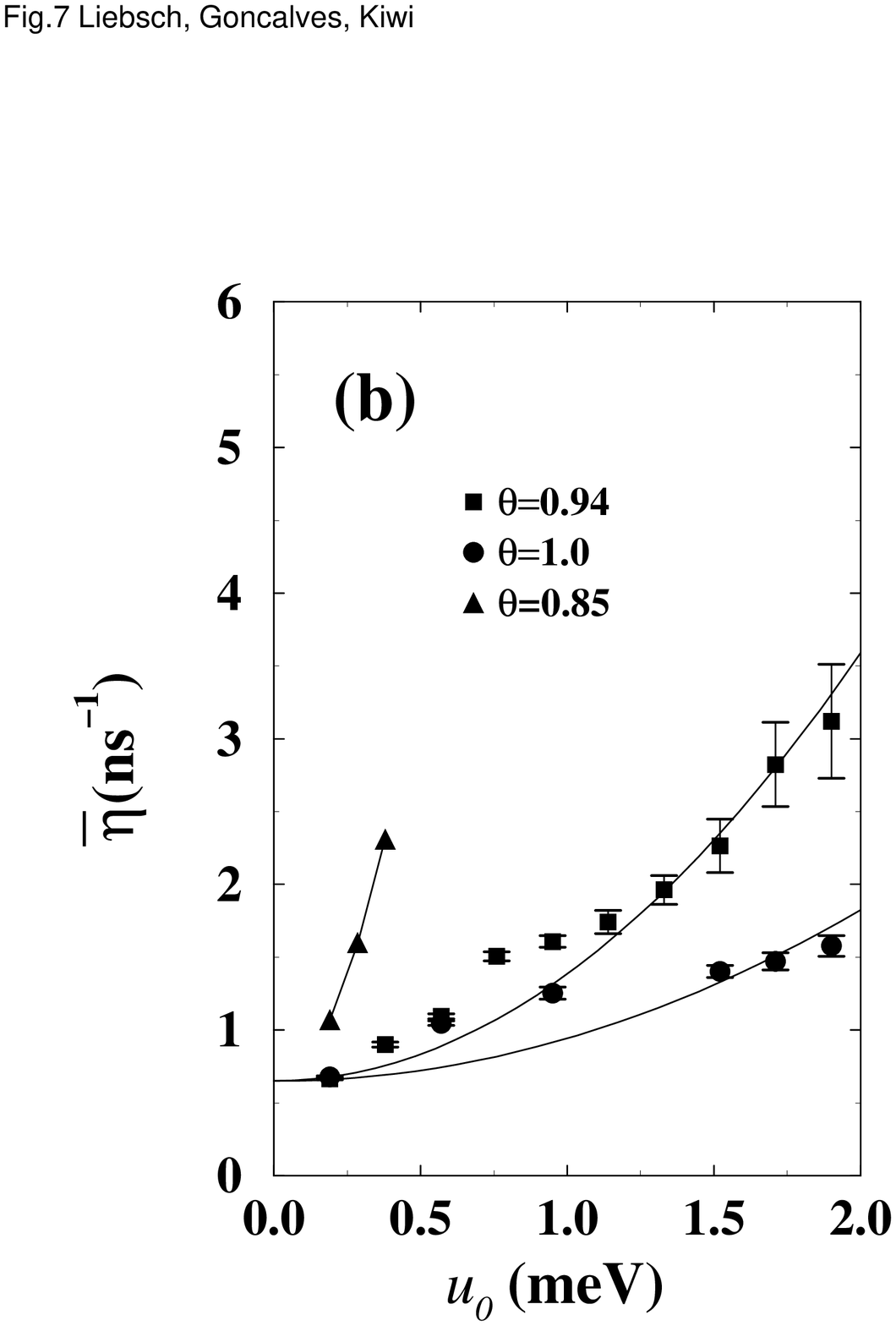}
\narrowtext
\caption{Net sliding friction $\bar\eta$ of Xe layers
on Ag as a function of corrugation amplitude $u_0$ for coverages
$\Theta=0.85, \ 0.94,\ 1.0$. (a) (100) surface; (b) (111) surface.
$\eta_\Vert = 0.65$~ns$^{-1}$.}
\end{figure}

Concerning the overall variation of $\eta_{ph}$ in the coverage
range $\Theta=0.85\ldots1.0$, on Ag(100) for $u_0=0.95$~meV the
variation is about 1~ns$^{-1}$, while for $u_0=0.5$~meV this
variation diminishes to less than $0.3$~ns$^{-1}$, and for
$u_0=1.4$~meV it is larger than 2~ns$^{-1}$. On Ag(111), in the
same coverage range, $u_0=0.3$~meV yields a variation of
$\eta_{ph}$ of about 1~ns$^{-1}$, whereas $u_0=0.2$~meV
($u_0=0.4$~meV) lead to much smaller (larger) variations.
Therefore $\eta_{ph}$ is sensitive to the substrate topology.

In view of the great sensitivity of $\eta_{ph}(\Theta)$ with
respect to the corrugation amplitude, it would be of interest to
have more accurate independent experimental and/or theoretical
determinations of the substrate/adsorbate potential.  In
principle, the corrugation amplitude may be estimated from the
lateral vibrational frequency $\omega_\Vert$ of adsorbed Xe,
which can be detected using inelastic He scattering. For Xe on
Cu(111), one finds \,$\omega_\Vert\approx
3\pm1$~cm$^{-1}$~\cite{braun}. Using the relation
\,$u_0=m\omega_\Vert^2/k^2$, one obtains values of $u_0$ in the
approximate range between 0.5 and 2~meV. From Eq.~(\ref{c}) it
follows that $\eta_{ph}$ varies with the fourth power of
$\omega_\Vert$.  This is born out by our simulations which show
that such barriers yield a phonon-induced friction in a very
wide range, \,$\eta_{ph}\approx 0.1\ldots5.0$~ns$^{-1}$, if the
coverage is varied between 1.0 and 0.85.

\subsection{Comparison with Experiment}
As pointed out in Section I, the quartz-crystal microbalance
measurements by Krim {\it et al.}~\cite{krim1} for thin Xe films on
Ag(111) show that the slip time \,$\tau=1/\bar\eta$\, exhibits a
characteristic dependence on Xe coverage (see Fig.~1).
Close to one monolayer, the slip time reaches a minimum of about
\,$\tau\approx 0.8\,$~ns and then increases to about 2.0~ns within a
coverage range of less than 0.3~monolayers. In this section we focus on
this coverage range (roughly the ``compression region'') since it
provides the most challenging aspect of the data.
The interpretation of the slip time at submonolayer
coverages in terms of simulations based on small repeated cells is
less reliable because they do not account for the formation of islands.
Also, the presence of defects and steps which are not included in
the theory then plays a larger role. On the other hand, beyond full
monolayer coverage, the observed slip time shows a much weaker coverage
dependence related to additional phonon dissipation in the partly
filled second monolayer.

In Ref. \cite{TS97} it is stated that
the minimum of $\tau$ corresponds to the uncompressed monolayer at
\,$n_a\approx0.0563$~\AA$^{-2}$ and that the subsequent steep rise
reflects the suppression of intra-adsorbate phonon excitations as the
Xe layer becomes more compressed (as noted above, the compressed
phase is only 6~\% denser than the uncompressed layer~\cite{dai94}.)
This coverage assignment implies that the slip time increases
(i)~when the coverage is reduced below that of the uncompressed
monolayer; and (ii)~when the coverage is increased beyond that of the
compressed layer. However, both of these implications are
implausible since in general incommensurate solid layers slide more
easily than fluids~\cite{cieplak94,cieplak94b,PN96}. Thus, the slip time should
decrease rather than increase (i)~when the coverage is lowered below that
of the uncompressed layer and (ii)~when it is increased beyond that of the
compressed layer, i.e., when the second layer is beginning to adsorb.
Since absolute coverage calibration in the experiment is difficult,
there exists a clearly appreciable uncertainty concerning the measured
sliding friction of a monolayer. In fact, it would appear more
plausible to associate the minimum of the slip time not with the
uncompressed layer but with a slightly lower coverage. Part of the
observed steep rise of $\tau$ would then correspond to the reduction
of intra-overlayer phonon processes upon reaching the uncompressed
layer coverage, and the remaining rise should reflect the
additional phonon reduction due to compression. The maximum of the observed slip
time should approximately correspond to the fully compressed phase.
According to these arguments, the uncompressed monolayer might have
$\tau\approx 1.3\ldots1.6$~ns, giving \,$\bar\eta\approx0.6\ldots
0.8$~ns$^{-1}$.

In Fig.~8 the phonon friction $\eta_{ph}(\Theta)$ is plotted as
a function of coverage, in the range
$\Theta=0.85\ldots1.0$,\ for several values of the corrugation
amplitude $u_0$.  For ease of comparison with the experimental
results, we include in Fig.~8(b) the measured $\bar\eta=1/\tau$
as a function of coverage in the region near one monolayer where
$\tau$ exhibits the characteristic positive slope.  To account
for the uncertainty of the measured coverage, and for the
expected behavior of the sliding friction as discussed above, a
shift of the data to lower coverages, by about
$0.1\ldots0.15$~monolayer, should be allowed for. 

\begin{figure}
\includegraphics[width=7.cm, clip=true]{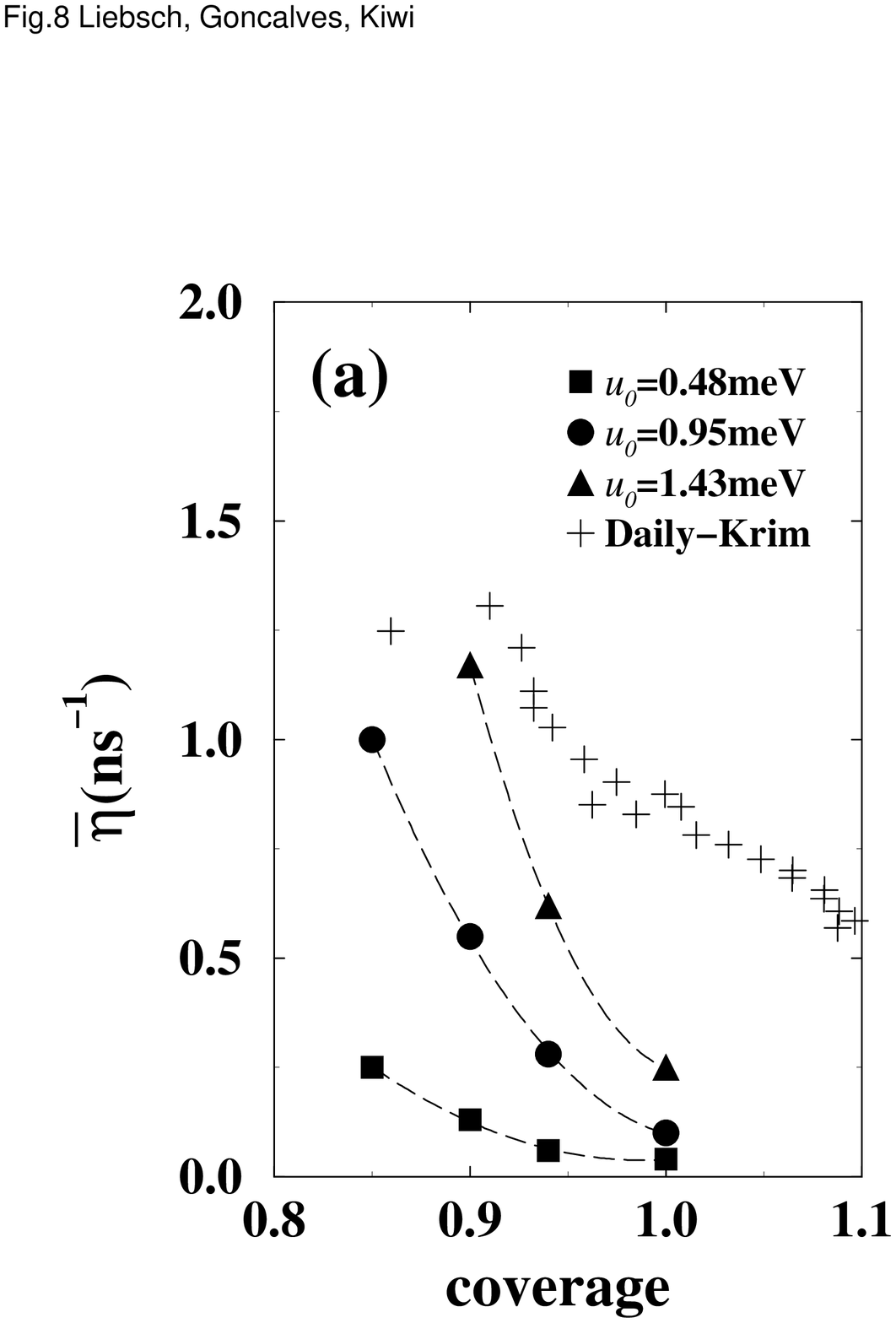}

\includegraphics[width=7.cm, clip=true]{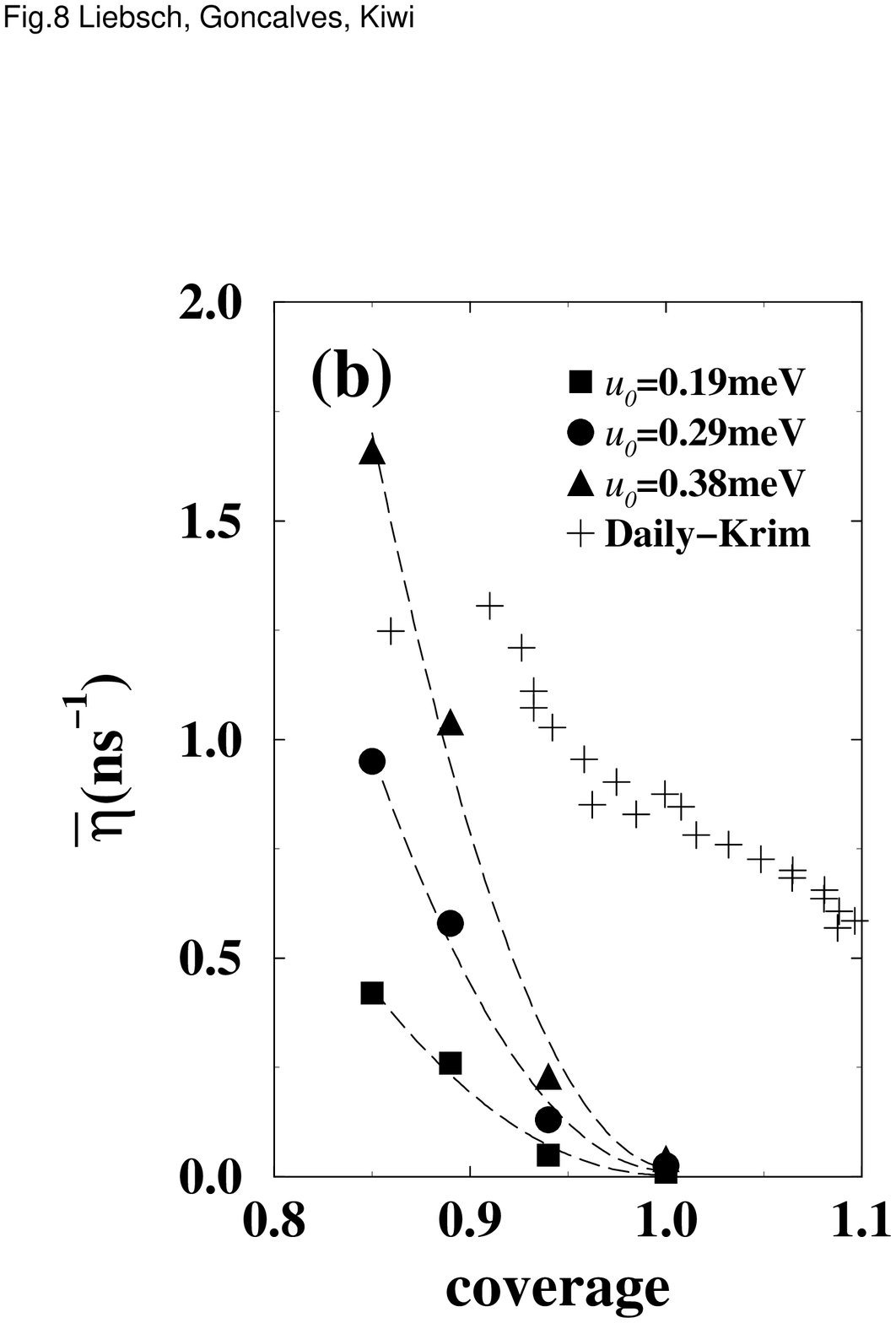}
\narrowtext
\caption{Calculated phonon friction $\eta_{ph}(\Theta)$
of Xe on Ag as a function of coverage $\Theta$ for several corrugation
amplitudes $u_0$. (a) (100) surface; (b) (111) surface. The dashed lines
are guides to the eye. The error bars indicate the uncertainty due to
fluctuations. Panel (b) also shows the measured sliding friction
$\bar\eta=1/\tau$ in the compression region where the slip time $\tau$
exhibits the characteristic positive slope (see Fig.~1).}
\end{figure}

The coverage
dependence of $\bar\eta$ in the range shown can be characterized
by two important features: the overall magnitude and the slope.
The comparison of $\bar\eta$ with the calculated results shows
that none of the functions $\eta_{ph}(\Theta)$ obtained for
different corrugation amplitudes $u_0$ provides an adequate
description of the data. However, for $u_0=0.19$~meV, the
calculated $\eta_{ph}$ has roughly the right slope even though
its magnitude at full monolayer coverage is far too small. If
$u_0$ is increased to $0.3\ldots0.4$~meV the slope of
$\eta_{ph}(\Theta)$ becomes rapidly larger than that of the
measured $\bar\eta$ while the magnitude near monolayer coverage
is still too small.  Only when $u_0$ is increased to much larger
values does $\eta_{ph}$, in the compressed phase, match the observed
$\bar\eta$, but the slope of $\eta_{ph}(\Theta)$ then is much
larger than observed.

As shown in Fig.~8(a), a similar picture is found on Ag(100). Although
the detailed dependence of $\eta_{ph}(\Theta)$ on the corrugation differs
appreciably from the one obtained for the (111) face, near monolayer
coverage, \,$\eta_{ph}$ remains less than $0.1$~ns$^{-1}$ as long as
\,$u_0<1$~meV, with a moderate slope for $\Theta<1$. Only if $u_0$
is increased to \,2~meV we find \,$\eta_{ph}\approx 0.5$~ns$^{-1}$
in the compressed phase. The slope of $\eta_{ph}(\Theta)$, however,
then quickly becomes extremely large.

Thus, our results imply that it is not possible to simultaneously
describe the magnitude {\it and} slope of the observed variation
of $\bar\eta(\Theta)$ exclusively in terms of phonon dissipation.
Only by adding an electronic friction contribution (which in the
narrow range shown in Fig.~8 can be assumed to be independent of
coverage) does the sum \,$\eta_{el}+\eta_{ph}(\Theta)$\, reproduce
the measured variation of $\bar\eta(\Theta)$. The comparison with
the data suggests that \,$\eta_{el}\approx0.5$~ns$^{-1}$ together
with \,$u_0\approx0.2$~meV (total barrier height $\approx0.9$~meV)
provides a qualitative description of the sliding friction measured 
on Ag(111) ---if one allows a shift of the data of 0.14~monolayer 
to lower coverages, of course. As discussed in Section
II, an overall single-atom electronic friction coefficient of this
magnitude is in agreement with surface resistivity measurements and
independent theoretical estimates. 

Interestingly, our findings for Xe/Ag differ from those on
Kr/Au.  As recently shown by Robbins and Krim \cite{RK98}, the
Kr/Au QCM data up to monolayer coverage can be fitted rather
well assuming negligible electronic friction. A finite
electronic friction would require a correspondingly smaller
substrate corrugation $u_0$ which, in turn, would yield too low
values of the phonon friction at submonolayer coverages.  We
note, however, that the analysis of the low-coverage data is
uncertain since the simulations are limited to repeated finite
cells and do not include the effect of defects or steps, nor do
they properly describe the formation of islands which most
likely occur under experimental conditions. Thus, if one
focusses on the slip time in the less problematic region near
monolayer coverage, i.e., roughly \,$0.85\leq\Theta\leq 1$, the
analysis in Ref.~\cite{RK98} indicates that the inclusion of a
finite electronic friction and a slightly reduced corrugation
$u_0$ is not incompatible with the data.

In addition, we point out that electronic friction of Kr on Au indeed
ought to be weaker than for Xe on Ag for the following reasons. Since
the Au $d$ bands lie only 2~eV below the Fermi level, in contrast to 4~eV
for Ag, the larger $s-d$ hybridization diminishes the polarizability of
the $s$ electron tails decaying into the vacuum region and thereby
reduces the probability of exciting substrate electron--hole pairs
via adsorbed particles. The larger work function of Au (5.3~eV compared
to 4.8~eV) also contributes to a stiffer, less polarizable electronic
density tail \cite{liebsch97,liebsch97b}. Moreover, since the atomic
polarizability of Kr is about 40\,\% smaller than that of Xe \cite{ZS},
the leading van der Waals part of the friction coefficient should
be much smaller than for Xe on Ag. For the same physical reasons,
the contributions associated with the Pauli repulsion and the
covalency of the adatom bond should also be smaller. The overall
electronic coupling of Kr on Au should therefore be significantly
weaker than for Xe on Ag. Thus, great caution is required when
comparing electronic excitation processes on different adsorption
systems ---even in the case of rare gas atoms on noble metal surfaces.

The comparison of Figs.~8(a) and (b) suggests that, for the same
corrugation amplitude $u_0$, Ag(111) yields a larger phonon friction
than Ag(100). This trend seems at first surprising since the
close-packed (111) face is generally regarded as the smoothest of
the low-index faces of an fcc crystal. On the other hand, as pointed
out in Section III, the potential barrier between neighboring hollow
sites is $2u_0$ for the (100) face and $0.5u_0$ for the (111) face,
while the total barrier between hollow sites across the substrate
lattice sites is $4u_0$ for the (100) face and $4.5u_0$ for (111).
Thus, the (111) surface has sharper potential maxima than the (100)
face. Since atoms of an incommensurate overlayer sample not only
the potential minima but the entire substrate surface, it is indeed
plausible that Ag(111) causes more Xe intra layer phonon friction than
Ag(100). According to this, the topology of the substrate is a relevant
feature if one want to make a comparison with the experiments.

In their analysis of an uncompressed Xe layer on Ag(100), Persson
and Nitzan~\cite{PN96} used $u_0=0.95$~meV and obtained
$\eta_{ph}\approx0.01$~ns$^{-1}$ (for \,$\eta_\Vert=0.62$~ns$^{-1}$,
they find $\eta_\Vert/\bar\eta =0.98\pm0.04$). From this result, they
concluded that the net friction at monolayer coverage is mainly of
electronic origin.  Instead, for the same parameters, we find
\,$\eta_{ph}\approx 0.3$~ns$^{-1}$, suggesting a phonon friction
roughly half as large as the electronic contribution.

On the other hand, Tomassone {\it et al.}\cite{TS97} assumed a total
barrier height of 2~meV on Ag(111), implying in our notation
\,$u_0=0.45$~meV. On the basis of the simulations it was concluded
that phonon friction dominates over possible electronic contributions.
For an uncompressed Xe layer, these authors obtained a phonon friction
\,$\eta_{ph}\approx1.8$~ns$^{-1}$, which is considerably larger than
what we find for the same parameters. According to Fig.~2 of Ref.
\cite{TS97}, in the compression region, the calculated slip time for
\,$u_0=0.45$~meV increases extremely rapidly in agreement with our
results. This increase, however, is much steeper than what is seen
in the experiment. Thus, a weaker corrugation, combined with a finite
electronic friction might, in fact, provide a better fit of the data
in this coverage range. The analysis of the submonolayer data is
again more complicated for the reasons discussed earlier.

\section{Conclusion}
We have performed molecular dynamics simulations for Xe layers on
Ag(100) and Ag(111) in order to determine the net sliding friction
in the presence of a constant lateral external force. We find that
$\bar\eta$ depends sensitively on the corrugation amplitude $u_0$
of the Xe/Ag interaction potential and on the single-adatom
microscopic friction parameter $\eta_\Vert$. For coverages approaching
one monolayer, the phonon-induced contribution $\eta_{ph}(\Theta)$
decreases rapidly, in agreement with the strong positive slope of the
observed slip time $\tau$. A detailed study of $\eta_{ph}(\Theta)$
as a function of corrugation amplitude $u_0$ suggests that, close to
monolayer coverage, it is not possible to consistently describe both
the magnitude {\it and} slope of the measured $\bar\eta=1/\tau$
without including a constant lateral electronic friction $\eta_{el}$.
Our simulations indicate that this electronic contribution should be
about 0.5 ns$^{-1}$ which is in qualitative agreement
with surface resistivity measurements and independent theoretical
estimates.

The overall picture that
seems to be consistent with the available information is that the
rapidly varying part of the observed sliding friction is due to
excitation of atomic vibrations within the overlayer and that
single-site electronic friction processes contribute a roughly
coverage-independent background. For a fully compressed Xe layer,
phonon dissipation is weaker than the electronic friction. As
the coverage is reduced, however, phonon processes rapidly become
important and eventually, near the minimum of the observed slip
time, dominate over the electronic dissipation channel.

In summary, even though both experimental and theoretical
results on sliding friction of Xe on Ag have inherent
uncertainties, our analysis indicates that both mechanisms
---phononic and electronic dissipation--- do contribute. The
relative weight of these channels depends strongly on coverage
in the crucial region near a full monolayer.

\acknowledgments
One of the authors (A.L.) gratefully acknowledges
the hospitality of the Facultad de F\'\i sica, Universidad
Cat\'olica, Chile, where this work was partially carried out, as well
as financial support by the European Community. A. L. also thanks
Prof. Abe Nitzan for making available his molecular dynamics code and
Dr. P. Ballone for several useful suggestions concerning the simulations.
M. K. was partially supported by the {\it Fondo Nacional de
Investigaciones Cient\'\i ficas y Tecnol\'ogicas} (FONDECYT, Chile)
under grant \#~1971212 and  S. G. under grant \#~3950028.

\noindent
$^*$e-mails:
a.liebsch@fz-juelich.de, \ sgonc@if.ufrgs.br, \ mkiwi@puc.cl

\end{multicols}

\begin{references}
\bibitem{bowden}  F. P. Bowden and D. Tabor, {\it Friction and
Lubrication}  (Methuen, London, 1967).

\bibitem{persson} B. N. J. Persson, {\it The Physics of Sliding
Friction}     (Springer, Heidelberg, 1997).

\bibitem{krim1}
E. T. Watts, J. Krim, and A. Widom, Phys. Rev. B {\bf 41}, 3466
(1990); A. Widom and J. Krim, Phys. Rev. E {\bf 49}, 4154 (1994); J.
Krim, D. H. Solina, and R. Chiarello, Phys. Rev. Lett. {\bf 66}, 181
(1991).

\bibitem{krim2}
C. Daly and J. Krim, Phys. Rev. Lett. {\bf 76}, 803 (1996); A. Dayo
and J. Krim, Proc. 8$^{th}$ International Conference on Vibrations at
Surfaces, Birmingham, England, June, 1996, Surf.  Sci. in press.

\bibitem{holzapfel}
C. Holzapfel, F. Stubenrauch, D. Schumacher, and A. Otto, Thin Solid
Films {\bf 188}, 7 (1990).

\bibitem{SH81} W. L. Schaich and J. Harris, J. Phys. C {\bf 11}, 65
(1981).

\bibitem{Sols82} F. Sols and F. Flores, Sol. St. Commun. {\bf 42},
687 (1982); F. Sols, F. Flores, and N. Garc\'\i a,  Surf. Sci.
{\bf 137}, 167 (1984).

\bibitem{P91} B. N. J. Persson, Phys. Rev. B {\bf 44}, 2177 (1991).

\bibitem{PV95}
B. N. J. Persson and A. J. Volokitin, J. Chem. Phys. {\bf 103}, 8679
(1995).

\bibitem{sokoloff95} J. B. Sokoloff, Phys. Rev. B {\bf 52}, 5318
(1995).

\bibitem{liebsch97} A. Liebsch, Phys. Rev. B {\bf 55}, 13263
(1997).

\bibitem{liebsch97b} A. Liebsch, {\it Electronic Excitations at
Metal Surfaces} (Plenum, New York, 1997).

\bibitem{cieplak94} M. Cieplak, E. D. Smith, and M. O. Robbins,
Science {\bf 265}, 1209 (1994).

\bibitem{cieplak94b} M. Cieplak, E. D. Smith, and M. O. Robbins,
Phys. Rev. B {\bf 54}, 8252 (1996).

\bibitem{PN96}
B. N. J. Persson and A. Nitzan, Surf. Sci. {\bf 367}, 261 (1996).

\bibitem{TS97}
M. S. Tomassone, J. B. Sokoloff, A. Widom, and J. Krim, Phys. Rev.
Lett. {\bf 79}, 4798 (1997).

\bibitem{inglesfield}
G. C. Aers and J. E. Inglesfield, Surf. Sci. {\bf 217}, 367 (1989).

\bibitem{steele}  W. Steele, Surf. Sci. {\bf 36}, 317 (1973).

\bibitem{tully} J. C. Tully, G. H. Gilmer, and M. Shugard, J. Chem.
Phys.  {\bf 71}, 1630 (1979).

\bibitem{dai94} P. Dai, T. Angot, S. N. Ehrlich, S. K. Wang, and H. Taub,
Phys. Rev. Lett.  {\bf 72}, 685  (1994).

\bibitem{braun} J. Braun, D. Fuhrmann, A. Siber, B. Gumhalter, and
Ch. W\"oll, Phys. Rev. Lett.  {\bf 80}, 125  (1998).

\bibitem{RK98} M. O. Robbins and J. Krim, Materials Research Society
Bulletin {\bf xxx}, 23 (1998).

\bibitem{ZS} A. Zangwill and P. Soven, Phys. Rev. A {\bf 21}, 1561 (1980).
\end{references}
\end{document}